\documentclass[conference,letterpaper]{IEEEtran}
\usepackage[letterpaper, left=0.7in, right=0.7in, bottom=0.95in, top=0.7in]{geometry}
\IEEEoverridecommandlockouts

\usepackage[font=footnotesize,caption=false]{subfig} 

\usepackage{cite}
\usepackage{amsmath,amssymb,amsfonts, mathtools, amsthm}
\usepackage{algorithmic}
\usepackage{graphicx}
\usepackage{textcomp}
\usepackage{xcolor}
\usepackage{booktabs}
\usepackage{glossaries}
\usepackage{subfig}
\usepackage{algorithm}
\usepackage{algorithmic}
\usepackage{bbm}
\usepackage{bm}
\usepackage{multirow}

\usepackage[normalem]{ulem}

\newacronym{bcd}{BCD}{Block Coordinate Descent}
\newacronym{leo}{LEO}{Low Earth Orbit}
\newacronym{isl}{ISL}{Inter-Satellite Link}
\newacronym{gsd}{GSD}{ground sample distance}
\newacronym{fov}{FoV}{field of view}
\newacronym{gtfp}{GTFP}{ground track frame period}
\newacronym{gs}{GS}{ground station}
\newacronym{fso}{FSO}{free-space optical}
\newacronym{smec}{SMEC}{satellite mobile edge computing}
\newacronym{mec}{MEC}{mobile edge computing}
\newacronym{pdd}{PDD}{Penalty Dual Decomposition}
\newacronym{aodv}{AODV}{Ad hoc On-demand Distance Vector}
\newacronym{cnn}{CNN}{convolutional neural networks}
\newacronym{dl}{DL}{Deep Learning}
\newacronym{dod}{DoD}{depth of discharge}
\newacronym{dqn}{DQN}{deep Q-learning}
\newacronym{gsl}{GSL}{ground-to-satellite link}
\newacronym{bm}{BM}{benchmark}
\newacronym{ml}{ML}{Machine Learning}
\newacronym{mdp}{MDP}{Markov decision process}
\newacronym{ngeo}{NGEO}{Non-geostationary orbit}
\newacronym{olsr}{OLSR}{optimized link state routing protocol}
\newacronym{ospf}{OSPF}{Open Shortest Path First}
\newacronym{pan}{PAN}{Path-Aware Networking}
\newacronym{qos}{QoS}{Quality of Service}
\newacronym{rl}{RL}{Reinforcement Learning}
\newacronym{drl}{DRL}{Deep \gls{rl}}
\newacronym{dnn}{DNN}{Deep Neural Network}
\newacronym{dql}{DQL}{Deep Q-learning}
\newacronym{e2e}{E2E}{end-to-end}
\newacronym{bgp}{BGP}{Border Gateway Protocol}
\newacronym{ibgp}{iBGP}{interior Border Gateway Protocol}
\newacronym{ebgp}{eBGP}{exterior Border Gateway Protocol}
\newacronym{as}{AS}{Autonomous System}
\newacronym{relu}{ReLu}{Rectified Linear Unit}
\newacronym{cdf}{CDF}{Cumulative Distribution Function}
\newacronym{ntn}{NTN}{Non-Terrestrial Networks}
\newacronym{lsatc}{LSatC}{\gls{leo} Satellite Constellation}
\newacronym{ai}{AI}{Artifical Intelligence}
\newacronym{ip}{IP}{Internet Protocol}
\newacronym{ue}{UE}{User Equipment}
\newacronym{pomdp}{POMDP}{Partially Observable Markov Decision Problem}
\newacronym{hol}{HOL}{Head of Line}
\newacronym{fifo}{FIFO}{First-In First-Out}
\newacronym{snr}{SNR}{Signal-to-Noise Ratio}

\newcommand{\ilm}[1]{\textcolor{black}{#1}}

\DeclareMathOperator*{\argmax}{arg\,max}

\def\BibTeX{{\rm B\kern-.05em{\sc i\kern-.025em b}\kern-.08em
    T\kern-.1667em\lower.7ex\hbox{E}\kern-.125emX}}

\title{Q-learning for distributed routing in LEO satellite constellations\vspace{-0em}}

\author{\IEEEauthorblockN{Beatriz Soret\IEEEauthorrefmark{1}, Israel Leyva-Mayorga\IEEEauthorrefmark{2}, Federico Lozano-Cuadra\IEEEauthorrefmark{1}, and Mathias D. Thorsager\IEEEauthorrefmark{2}}
\IEEEauthorblockA{\IEEEauthorrefmark{1}Telecommunications Research Institute, University of Malaga, Spain (bsoret@ic.uma.es, fedeloz@uma.es)\\
\IEEEauthorrefmark{2}Department of Electronic Systems, Aalborg University, Denmark (ilm@es.aau.dk, mthors18@student.aau.dk)  
}} 
\date{}

\def\subparagraph{} 
\usepackage{titlesec}
\titlespacing*{\section}{0pt}{*1}{*1}
\titlespacing{\subsection}{0pt}{*1}{*1}

\renewcommand{\thesubsubsection}{\arabic{subsubsection}}

\titleformat{\subsubsection}[runin]{\itshape}{\thesubsubsection)}{1em}{}
\titlespacing*{\subsubsection}{\parindent}{0pt}{*1}

\begin{document}

\bstctlcite{IEEEexample:BSTcontrol}

\maketitle
\begin{abstract}
End-to-end routing in \glspl{lsatc} is a complex and dynamic problem. The topology, of finite size, is dynamic and predictable, the traffic from/to Earth and transiting the space segment is highly imbalanced, and the delay is dominated by the propagation time in non-congested routes  and by the queueing time at \glspl{isl} in congested routes. Traditional routing algorithms depend on excessive communication with ground or other satellites, and oversimplify the characterization of the path links towards the destination. We model the problem as a multi-agent Partially Observable Markov Decision Problem (POMDP) where the nodes (i.e., the satellites) interact only with nearby nodes. We propose a distributed Q-learning solution that leverages on the knowledge of the neighbours and the correlation of the routing decisions of each node. We compare our results to two centralized algorithms based on the shortest path: one aiming at using the highest data rate links and a second genie algorithm that knows the instantaneous queueing delays at all satellites. The results of our proposal are positive on every front: (1) it experiences delays that are comparable to the benchmarks in steady-state conditions; (2) it increases the supported traffic load without congestion; and (3) it can be easily implemented in a \gls{lsatc} as it does not depend on the ground segment and minimizes the signaling overhead among satellites.
\end{abstract}
\vspace{-.1cm}
\glsresetall
\section{Introduction}
\glspl{lsatc}, with hundreds or even thousands of satellites working all together as a communication network, are one of the pillars to attain global connectivity~\cite{Leyva-Mayorga2020} by extending cellular coverage, serving as a global backbone, and offloading congested terrestrial infrastructure, as well as supporting new remote inference and intelligent applications. 
Realizing this vision will require technical solutions for an efficient use of the scarce wireless and computing resources on space.

At the network level, a packet-based network requires a routing algorithm for deciding the directions which must be used to reach the destination. To make these decisions, the network is modeled as a graph where the network nodes are the transmitters and receivers and the edges represent the links between nodes. The weights of the edges are defined as a function of the target network performance metric. Hence, the routing problem is that of finding an optimal path, defined as a sequence of vertices or edges, to forward the packets from source to destination that minimize the sum of the weights.
In wired terrestrial networks, the links of the routers are fixed, have a mostly stable capacity, and do not suffer from interference, so the routing problem can be readily solved with well-known Dijkstra’s shortest path algorithm~\cite{dijkstra1959}. 
On the other hand, the underlying graph in wireless networks is time-varying, i.e., prone to capacity degradation due to the movement of the terminals, to blockages, and to interference, and the routing algorithm must dynamically adapt to the changes in the network graph.



While there have been attempts to adapt routing algorithms for terrestrial wireless ad hoc networks to \glspl{lsatc}, these have not yet proven to be successful because \gls{lsatc}s presents distinctive characteristics that are not considered in their design. First, {the network topology is mostly symmetric} yet the satellites move rapidly with respect to the ground infrastructure and with respect to some other satellites. Because of the dynamic nature of the constellations, traditional routing protocols may suffer from excessive signaling overhead to maintain the routing tables, whereas routing protocols for ad hoc networks cannot exploit the symmetry of the constellation. Second, {the movement is fully predictable} once the ephemeris, that is, the orbital parameters of the satellites, is known and the size of the constellation is finite and well known. Because of this, the network topology can be easily calculated in advance to generate, at least partially, the routing tables to be used at a specific point in time. In fact, we have exploited this knowledge to develop simple and efficient algorithms to find establish the best \gls{isl} matching as part of such network discovery for \gls{lsatc}s~\cite{Leyva-Mayorga2021}, which then can be used as an input to the routing algorithm. Third, {the links connect satellites separated by long distances}. As a consequence, the propagation latency is a dominant factor in the total latency. This was observed in~\cite{Rabjerg2021}, where traditional unipath source routing lead to an end-to-end average routing latency of around $100$~ms, where the contribution of the propagation delay ranged from $37$\% to up to $66$\% on average, depending on the load level. Finally, the ground traffic injected to the constellation network is highly imbalanced and dependent on the connectivity to the ground segment, which requires a realistic model of the ground topology, too.

\begin{figure}[t]
\centering
{\includegraphics[width=3.5in]{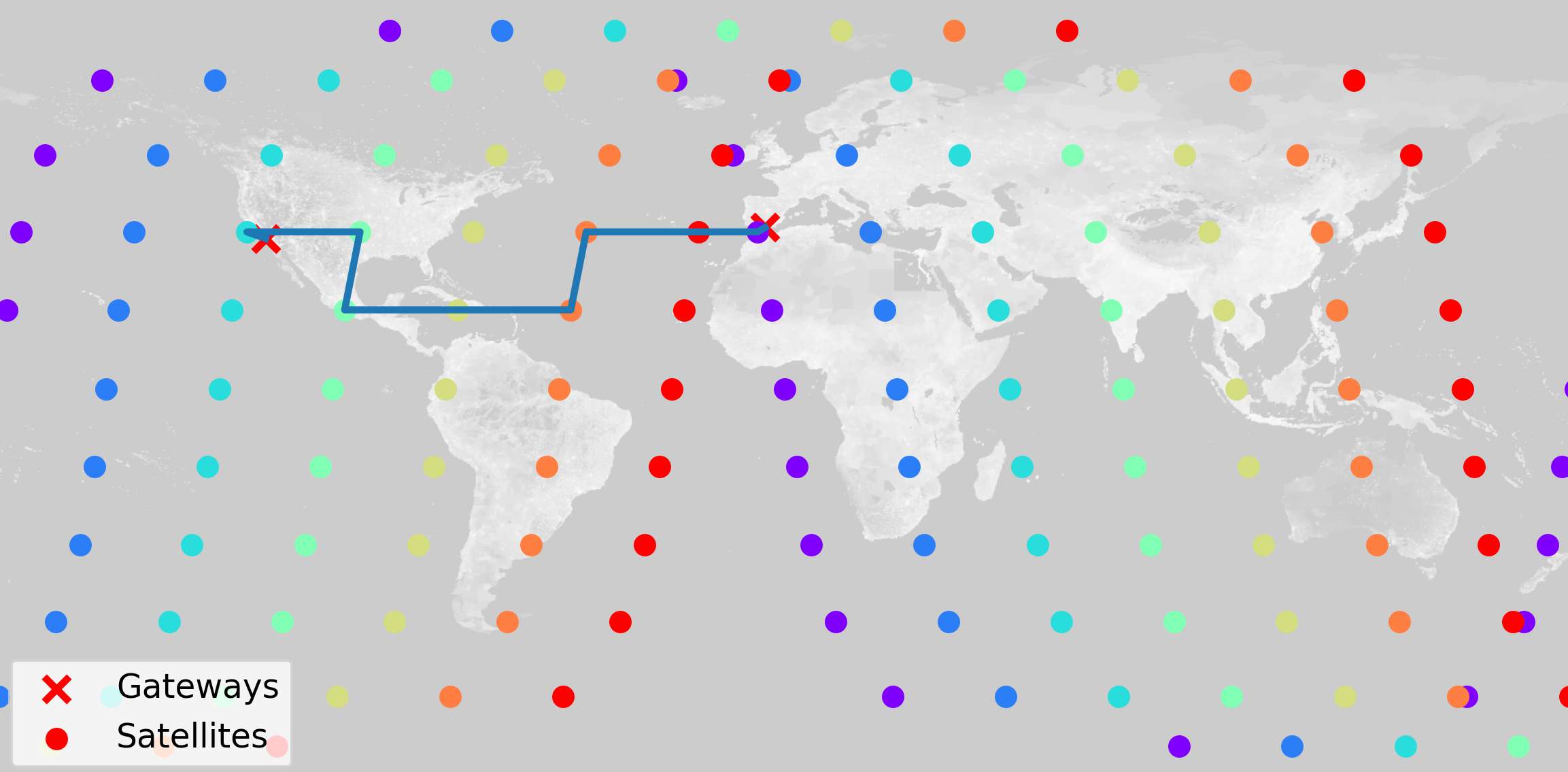}}
\caption{Example of a route connecting a gateway in Málaga and a gateway in Los Angeles.} 
\label{fig:map}\vspace{-0.5cm}
\end{figure}

This paper studies the problem of E2E routing between nodes in the ground segment, consisting of multiple gateways, through the \gls{lsatc} without relying on further ground infrastructure (see Fig.~\ref{fig:map}). We aim at developing a fully-distributed learning approach to achieve robust and low latency E2E communication while minimizing the signaling overhead. Previous works have employed different \gls{ml} algorithms~\cite{fadlullah2017deep}, including \gls{rl} and supervised learning. 
Deep \gls{rl} has been used to account for the battery ageing on the centralized routing decisions in~\cite{Liu2021}, but it is not clear whether the routing agent can find a stable solution when multiple routing agents interact with each other.
Even when utilizing deep learning models, it is challenging to account for the time-varying queueing time at the links. For example, \glspl{dnn} have been used for multi-agent multi-objective optimization by exploiting the optimal substructure of the routing problem: if a node knows the latency to communicate to each of its neighbors and the latency to communicate from these neighbors to the destination, it can readily compute the shortest path~\cite{Liu2022}. Therefore, the task of the \gls{dnn} is to estimate the latency from a satellite's neighbors to the destination. However, even in these cases and with this optimal substructure, the model of the queueing time is too simplistic, e.g., using truncated Gaussian distributions~\cite{Liu2022}. Our work differentiates from the state-of-the art, including the previous two studies~\cite{Liu2021, Liu2022}, in the sense that we consider the ground segment topology and its connectivity to the space segment, and solve the routing problem in a realistic setup with queues at each \gls{isl} and multiple simultaneous data flows that affect these queues.

In this paper, we have a comprehensive model of the \gls{lsatc} and ground infrastructure, based on which we formulate the routing as a \gls{pomdp}. Then, we propose a Q-learning algorithm where each satellite is an agent that collects only local information for a fully distributed solution. 
This learning has analogies with the Q-routing first proposed in \cite{qrouting1993boyan} for dynamically changing terrestrial networks and using only local communication at each node to keep accurate statistics on which routing decisions lead to minimal delivery times. 
We evaluate and compare our solution with two benchmark algorithms, one state-of-the-art shortest-path and a genie-aided scheme to minimize the latency. The results show the advantage of our Q-routing algorithm in terms of latency, supported load and, primarily, implementability.




\section{System Model} \label{sec:systemmodel}
The communication network is composed of the space segment and the ground segment. Although the system is dynamic, we omit the time index in the notation related to the graphs for the sake of simplicity.

\vspace{1in}
\noindent\textbf{Space segment.} The satellite communication network modeled as a graph consisting of a total of $N$ satellites evenly distributed in $M$ orbital planes. $\mathbb{S}$ denotes the finite set of satellite nodes and $\mathcal{E}$ the set of transmission links that are the edges of the graph. 
Each orbital plane $m\in\{1,2,\dotsc,M\}$ is deployed at a given altitude $h_m$~km above the Earth's surface, at a given longitude $\epsilon_m$~radians, has a given inclination $\delta$, and consists of $N_m=N/M$ evenly-spaced satellites. 
The \gls{leo} satellites are equipped with one antenna for ground-to-satellite communication and four antennas for inter-satellite communication. Two of the latter antennas are located at both sides of the roll axis (i.e., front and back of the satellite) and are used to communicate with immediate neighbors in the same orbital plane (intra-plane \gls{isl}). The other two antennas are located at both sides of the pitch axis and are used to communicate with satellites in different orbital planes through (inter-plane \gls{isl}). We denote $\mathcal{E}_i$ the set of feasible edges of satellite $i$, i.e., the \gls{isl} that are currently available for communication. $\mathcal{E}_i$ has a maximum size of four: $|\mathcal{E}_i| \leq 4$. $\mathcal{E}_i$ is dynamically updated using the algorithm in~\cite{Leyva-Mayorga2021}. $\mathcal{E}_S$ is the set of edges of the space segment, i.e., $\mathcal{E}_S = \sum_{i\in \mathbb{S}} \mathcal{E}_i$.

\noindent\textbf{Ground segment.} 
Let $\mathbb{G}$ to be the set of gateways distributed across the Earth in relevant locations, denoted as $\left(g_\text{lat},g_\text{lon}\right)$ for $g\in\mathbb{G}$. The mobile devices at ground level communicate with the constellation through these gateways, which 1) generate relatively large packets that are transmitted to the nearest satellite (a.k.a. feeder link) by collecting data from the nearby users and/or data centers and 2) collect the data transmitted by other gateways through the \gls{lsatc} to the users in the vicinity (a.k.a. service link). 

The gateways maintain one \gls{gsl} with their closest satellite at all times. These \gls{gsl}s constitute the set of edges $\mathcal{E}_G$, and $\mathcal{E}_{i_G}$ is the \gls{gsl} between satellite $i$ and the closest gateway, which takes the value null when there is none. With these definitions, we distinguish three cases for the link $ij$ between node $i$ and node $j$: (1) $i\in\mathbb{G}$ and $j\in\mathbb{S}$ corresponds to a feeder link: uplink from a gateway to a satellite, and $ij \in \mathcal{E}_G$; (2) $i\in\mathbb{S}$ and $j\in\mathbb{G}$ corresponds to a service link: downlink from a satellite to a gateway, and $ij \in \mathcal{E}_G$; (3) otherwise, both $i$ and $j$ are satellites and we have an \gls{isl}, and $ij \in \mathcal{E}_i \subset \mathcal{E}_S$. The slant range between nodes $i$ and $j$ is denoted as $||ij||$.

In both cases, \gls{isl} and \gls{gsl}, the data rate between nodes $i$ and $j$ is selected by using the highest modulation and coding scheme, among those defined for DVB-S2, that ensures reliable communication with the current \gls{snr}. That is, let $\left\{\rho\right\}$  be the set of spectral efficiencies $\rho$\,bits/s/Hz that can be achieved with the DVB-S2 modulation and coding schemes \cite{dvb_s2} and $\text{SNR}_\text{min}(R')$ be the minimum \gls{snr} to achieve reliable communication with $\rho$. We assume a free-space path loss model, so the data rate for communication from node $i$ to $j$ is set to
\begin{equation} \label{eq:rate}
    R(i,j) = W\max\left\{\rho:\frac{P_r(i,j)}{k_BT_SB}\geq \text{SNR}_\text{min}(\rho)\right\}, 
\end{equation}
\noindent where $W$ is the bandwidth, $P_r(i,j)$ is the received power at node $j$ from $i$ including the antenna gains and free-space path loss, $k_B$ is the Boltzmann's constant, and $T_S$ is the system noise temperature.

\noindent\textbf{Traffic generation.}
We consider a scenario with realistic packet generation, queueing, and transmission, where each active gateway transmits an equal amount of data among the rest of the gateways in $\mathbb{G}$ through the \gls{lsatc}, which then distribute the data among the users connected to it. The case in which a gateway generates data to the users directly connected to it is not considered as does not lead to the transmission of data through the \gls{lsatc}. Let $\lambda_\text{UL}^{(g)}$ be the uplink data generation (i.e., arrival) rate at gateway $g$ and \ilm{$\lambda^*$  be the maximum supported traffic load in the network, calculated from the uplink and downlink \gls{gsl} rates. Next, we define the total traffic load as  $\ell=\sum_{g\in\mathbb{G}}\lambda_\text{UL}^{(g)}/\lambda^*$. Building on these, the amount of data transmitted to a gateway is 
\begin{equation}    
\lambda_\text{DL}^{(g)}=\sum_{i\in\mathbb{G}\setminus \{g\}} \frac{\lambda_\text{UL}^{(i)}}{|\mathbb{G}|-1}=\frac{\ell\lambda^*-\lambda_\text{UL}^{(g)}}{|\mathbb{G}|-1}\leq R(i,g),
\end{equation}
where $R(i,g)$ is the downlink data rate for gateway $g$.}
Data is generated at the gateways following a Poisson distribution with rate $\lambda_\text{UL}^{(g)}$ and a block of $B$\,bits is generated when a source gateway has $B$ bits with the same destination gateway. 



\noindent\textbf{Routing.} The routing algorithm at each satellite aims at relying each received packet $p(d)$ towards the destination $d$. For this, each satellite has a transmitting buffer of maximum size $Q_{max}$ that follows a \gls{fifo} policy. If the buffer is not empty, the satellite takes the \gls{hol} data packet and delivers it to one of the neighbouring nodes. Packets arriving to a full buffer are discarded. The set of nodes visited by packet $p$ is denoted by $\mathcal{P}_p$.

\noindent \textbf{Latency.}
The  one-hop latency to transmit a packet $p$ of length $B$~bits from $i$ to $j$ can be calculated as follows. 
First, the \textit{queue time} at the transmission queue $t_q(i)$ is the time elapsed since the packet is ready to be transmitted until the beginning of its transmission. The transmitting buffer at satellite $i$ has a length $q_i$, a constant packet size $B$, and follows a \gls{fifo} policy. Although we consider a transmission from $i$ to $j$, the rest of packets in the queue can have different destinations and therefore use different links from the set $\mathcal{E}_i$ and $\mathcal{E}_{i_G}$, with different rates $R(i,\cdot)$. The queue time is hence given by $t_q(i) = \frac{q_i \cdot B}{R(i,\cdot)}$ where $R(i,\cdot)$ is calculated with equation (\ref{eq:rate}) and depends on the link used for the transmission of each en-queued data packet ahead $p$. Second, the \textit{transmission time}, which is the time it takes to transmit $B$~bits at $R(i,j)$~bps. Third, the \textit{propagation time}, which is the time it takes for the electromagnetic radiation to travel the distance $||ij||$ from $i$ to $j$. Hence, we have
\begin{equation} \label{eq:latency_metric}
    L(i,j) = \!\underbrace{ t_q(i) }_\text{Queue time}+ \!\underbrace{ \dfrac{ B }{ R(i,j) } }_\text{Transmission time}+\underbrace{ \dfrac{||ij||}{c} }_\text{Propagation time}.
\end{equation}

\begin{figure*}[t]
\centering
{\includegraphics[width=7in]{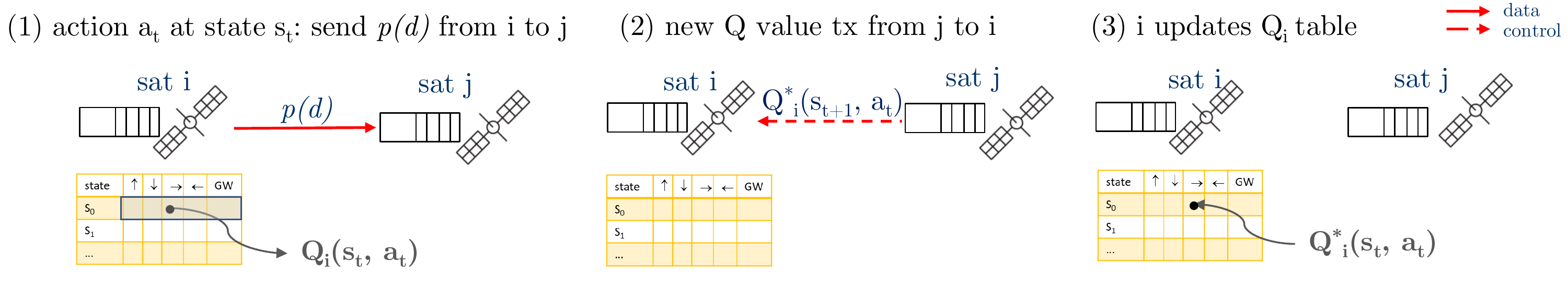}}
\caption{Dynamics of the Q-learning algorithm: Each satellite maintains its own Q-table. (1) Upon arrival of packet $p(d)$ at sat $i$, the best action $a_t$ is selected, i.e., forward the packet to sat $j$. (2) Sat $j$ calculates the new Q-value and sends it to sat $i$, who (3)  updates the Q-table accordingly. }
\label{fig:qrouting} \vspace{-0.5cm}
\end{figure*}

It was observed in~\cite{Rabjerg2021} that the propagation time is the dominating factor in a non-congested satellite constellation, whereas the queueing time grows with the system load. 

\noindent \textbf{Benchmarks.}
We compare the performance of our proposed intelligent Q-routing algorithm to two source routing approaches that use Dijkstra's algorithm to find the shortest path to the destination. In the first one, called the \emph{data rate \gls{bm}}, the weights of the edges are the inverse of the data rate, namely $w_{i,j}=1/R(i,j)$. This is a traditional routing approach that leads to choosing  routes with high data rate links. In the second one, called the \emph{latency genie \gls{bm}}, we assume instantaneous knowledge of the state of the queues at all satellites. By doing so, the source gateway is able to choose the route that minimizes the E2E latency based on the queueing time at the time a new packet is ready to be transmitted. Note that due to the long E2E latency, the latency genie \gls{bm} is not optimal, as the state of the queues at the satellites at the time the packet is received at each of them might be different to the one used to calculate the route.

\section{Learning framework} \label{sec:learning}
A \gls{lsatc} represents a networked multi-agent problem where agents are partially connected and interact uniquely with nearby agents. Although there is a global objective, we aim at a  fully-distributed solution where each satellite is an independent learning agent that learns its routing policy. 
Since each satellite only has a local observation of the environment founded on the local network state and the restricted communication with the neighbouring satellites, the decision-making problem is based on a partially observable state. 

The problem is formulated as a \gls{pomdp} with a 4-tuple $(\mathcal{S}, \mathcal{A}, P(s,a), \mathcal{R}(s,a))$ where the observation agents have a partial knowledge of the underlying system state, specifically obtained from their own queues and links and from the feedback of the neighbouring satellites. $\mathcal{S}$ is the state space, $\mathcal{A}$ is the action space, and $P(s,a)$ and $\mathcal{R}(s,a)$ are the probability of choosing action $a$ and its corresponding reward when in state $s$, respectively.

In Q-learning, the agent aims at finding out the optimal policy that maximizes the long-term accumulated reward. For this, we define the state-action value $Q$ to describe the expected reward of taking action $a_t$ under state $s_t$ at time $t$ from which the agent learns. Mathematically,
\begin{equation}\label{eq:q}
    Q(s_t, a_t) = (1-\alpha){Q(s_{t}, a_{t})} + \alpha \left( r_t + \gamma \underset{a}{\max} {Q(s_{t+1}, a)} \right),
\end{equation} 
\noindent where $\gamma$ is the discount factor that adjusts the importance of rewards over time. The goal is
to find out the optimum policy so that the total reward can
be maximized (or the total penalty can be minimized) over
a horizon of future time intervals given the current state of the agent. At each time instant $t$ the agent takes an action $a_t$, observes the reward $r_t$, and enters a new state $s_{t+1}$ that depends on the previous state $s_t$ and the selected action. The Q-table is updated in each step. At a given state, the algorithm compares the expected utility of all the available actions and chooses the action accordingly. 

\noindent\textbf{Time.} An episode is an instance to be solved by the learning algorithm. The episode is composed by $T$ steps, indexed by $t = 0, 1, ..., T$. The agent uses the environment’s rewards to learn, over time, the best action to take in a given state. In the notation, subindex $t$ to indicates the dependency upon the steps. 
At different time steps, the state observed by the agent is time varying due to the dynamic change of network traffic and the dynamic change of the topology. However, we note that these two time scales are very different. First, the satellite constellation moves following the laws of orbital motion, determined by the altitude of \gls{leo} satellites. The satellite passes for an observer at the Earth surface or between satellites in different orbital planes are at the scale of few minutes~\cite{Leyva-Mayorga2021}. Second, the time scale of the learning episodes is dictated by the arrival of packets, which is typically, much shorter
As observed in the results, the agents learn the new paths in less than $0.5$ s, therefore at a much faster pace than the movement of the constellation. Hence, we can model the system as an observation of the environment at specific time instants and keep the time index only for the learning episodes. 

\noindent\textbf{Environment.} The environment is the directed graph communication network described in Section~\ref{sec:systemmodel}.  

\noindent\textbf{State space.}
We denote the state space of agent $i$ as $S_i : \{L_i, N_i \}$, where $d_p$ is the destination of the current packet, $L_n$ is the local information at $i$ and $N_i$ is the information shared by neighbouring satellites to $i$. 
Specifically, $L_i$ is the information of the packet destination extracted from the packet header and the link connectivity $\mathcal{E}_i$, whereas $N_i$ contains the link quality and buffer congestion to the two intra-plane and two inter-plane neighbours each of them encoded with two bits: \mbox{$s_t = 2$} is reserved for the case with a long queue or unavailable link, and \mbox{$s_t = 0$} and \mbox{$s_t = 1$} reflect uncongested cases (empty/short queues) with high and low link capacity, respectively. This simple encoding minimizes the state space and alleviates the computation cost, which is an advantage for satellites with limited computation capabilities. In the future, we will extend the space space and apply other advanced learning techniques to characterize the tradeoff between complexity of the learning algorithm and performance gain.

\noindent\textbf{Reward.}
Let us assume a packet $p$ from source $s$ with destination $d$ arrives to satellite $i$. Satellite $i$ forwards the packet to satellite $j$. For this action, the immediate reward considers the two main contributors to the end-2-end delay: (1) the propagation time, for which the topology and the slant range of each decision should be considered; (2) the queueing time, which becomes dominant when the system gets congested~\cite{Rabjerg2021}. Moreover, extra rewards/penalties are given when reaching the destination and for avoiding loops. The reward for the action at satellite $i$ is defined as 
\begin{equation}
\label{eq:rewards}
    r_t = \begin{cases}
    r_{\text{del}} & jd \in \mathcal{E}_G \\
    r_{\text{loop}} &j \in \mathcal{P}_p
    \\
    r_\text{queue} + r_\text{dist} & \text{otherwise}    \end{cases}
\end{equation}

\begin{equation}
\label{eq:queue}
   r_\text{queue} = w_1 \cdot \left(1-10^{t_{q}(j)}\right) 
\end{equation}
\begin{equation}
\label{eq:dist}
r_\text{dist} = w_2 \cdot \frac{||id|| - ||jd|| + ||sd||}{||sd||}
\end{equation}

\noindent where $w_1, w_2$ are adjustment constants, $t_{q}(j)$ is the time spent in the queue of the next hop, i.e., satellite $j$; $||id||-||jd||$ is the slant range reduction of the decision, i.e., the distance difference between $j$ and $d$ and between $i$ and $d$; $||sd||$ is the total slant range between $i$ and $d$; and $r_\text{loop}$ and $r_\text{del}$ are extra penalties/rewards for loops and delivery, respectively. If the data block has been sent to a satellite whose linked gateway matches the destination of the data block, the agent will receive the extra reward $r_{\text{del}}$. To avoid loops, the penalty $r_{\text{loop}}$ is applied if the agent sends the data block to a satellite where it has already been. Otherwise, the first exponential term makes the penalty grow faster as the queue time increases. The second term in the sum, related to the slant range, quantifies and normalizes the distance towards the destination. 

\noindent\textbf{Action space.}
The action decision $a_t$ is the next hop $j$ selected from the set $\mathcal{E}_i \cup \mathcal{E}_{i_G}$, i.e., one of the neighbouring satellites or the link towards a gateway. 

\noindent \textbf{Q-routing algorithm.} 
The pseudo-code of the Q-routing algorithm is in \mbox{Algorithm 1}, and the dynamics of the routing and Q-tables updates in Fig. \ref{fig:qrouting}. The algorithm has the following characteristics:

\noindent\emph{$\varepsilon$-greedy policy}:
The agent explores with probability $\varepsilon$ new paths (i.e., selects a random action), and with probability $1-\varepsilon$ chooses the one that maximizes the expected reward. The value of $\varepsilon$ is high in the beginning, when the agent has not learnt the routes, and then it is exponentially decreased. 

\noindent\emph{Updating the Q-tables}:
In the conventional formulation of Q-learning, each agent updates its own q-table based on the new state and reward. However, this is not effective in our problem, because the actions taken by satellite $i$ are observable in the state change of the neighbouring satellites, more specifically in the increased queue length of the next hop $j$. Therefore, we modify the usual formulation to reflect this partial knowledge and correlation among actions. Mathematically, the new Q-value $Q^*$ is updated as
\begin{equation}\label{eq:q}
    Q^{*}_i(s_t, a_t) = (1-\alpha){Q_i(s_{t}, a_{t})} + \alpha \left( r_t + \gamma \underset{a}{\max} {Q_j(s_{t+1}, a)} \right)
\end{equation} 

\noindent\emph{Minimal feedback information}:
As illustrated in the steps of Fig.~\ref{fig:qrouting}, the algorithm minimizes the interaction with nearby satellites. Specifically, the Q-value is the only feedback after the successful reception of a packet, over a link that has been previously established.

\section{Results and discussion} \label{sec:results}

\begin{algorithm}[t]
 \caption{Q-routing algorithm.}
 \begin{algorithmic}[1]
 \renewcommand{\algorithmicrequire}{\textbf{Initialize:}}
 \REQUIRE Initialize $Q_i(s,a)\;\;\forall i$ arbitrarily

    \FOR { $t=1,2,\dotsc,T$}
        \STATE {Get next packet $p_i \in \mathbb{P}$ arriving at satellite $i$ with destination $d$}
        \STATE Update $\mathcal{E}_S$ and $\mathcal{E}_G$ using \cite{Leyva-Mayorga2021}
 \IF {$||id|| \in \mathcal{E}_{G}$}
\STATE Deliver $p_i$ to destination $d$
\STATE $r_t = r_\text{del}$
\ELSE
\IF {$r\sim U(0,1) < \varepsilon$}
\STATE Select a random action $a_t$
\ELSE
\STATE Select $a_t = {\argmax}_a Q_i(s_t, a)$

 \ENDIF
 \IF {$a_t = j \text{ AND } j\in \mathcal{P}_p$}
 \STATE $r_t = r_\text{loop}$
 \ELSE
 \STATE $r_t = r_\text{queue} + r_\text{dist}$
 \ENDIF
 \STATE $Q_i^{*}(s_t, a_t) \leftarrow \text{Eq.} (\ref{eq:q})$

 \ENDIF
    \ENDFOR
 \end{algorithmic}
 \label{al:param_searching} 
 \end{algorithm}%
Results were obtained by a simulator developed in Python. We consider a Kepler constellation with $M=7$ orbital planes at heights $h_m = 600$ km and $N_m = 20$ satellites per orbital plane. There are up to $18$ transmitting gateways at ground positions around the globe, mostly taken from the existing KSAT network\footnote{https://www.ksat.no/services/ground-station-services/}. Specifically, we locate the gateways in Malaga (Spain), Los Angeles (US), Aalborg (Denmark), Cordoba (Argentina), Tolhuin (Argentina), Inuvik (Canada), Nemea (Greece), Nuuk (Greenland),  Bangalore (India), Tokyo (Japan), Port Louis (Mauritius), 
Awarua (New Zealand), Svalbard (Norway), Vardø (Norway), Panama (Panama), Azores (Portugal), and
Singapore (Singapore). In the results, the simulator takes the first $|\mathbb{G}|$ from the sorted list above, with $2 \leq |\mathbb{G}| \leq 18$ \ilm{and traffic load $\ell=0.85$. We observed, through our simulations, that this range and traffic load allow us to analyze very low load up to scenarios with high congestion. } 
The communication parameters used for the simulations are as follows. The transmission power is $10$\,W for the satellites and $20$\,W for the gateways. The carrier frequencies are $20$\,GHz for downlink, $30$\,GHz for uplink, and $26$\,GHz for \gls{isl}. We consider parabolic antennas of diameter $33$\,cm at the gateways and of $26$\,cm at the satellites. The system bandwidth for all the links is $W=500$\,MHz. The packets length is $B=64.8$\,kbits.

We first do an analysis of the stability (congestion level) of each of the routing schemes, which determines the maximum traffic load supported by the algorithms. 
For this, we define that a path (i.e., source-destination pair) has stable E2E communication if it has reached a steady state where the E2E latency is not increasing with time. That is, an increase in E2E latency indicates that congestion is building up. To determine whether a route is stable, we perform a t-test based on the estimated slope $\hat{\beta}_1$ obtained from performing linear regression over the E2E latency of the packets in the route. Specifically, the independent variable is set to be the packet index in the route and the dependent variable is set to be the latency in ms for each packet. Then, we perform linear regression with the model $Y_i = \beta_0 +\beta_1X_i+\epsilon$ using the last $200$ packets received at the destination to avoid considering the training period. Then, we perform a t-test  with the null hypothesis $\text{H}_0\!: \beta_1\leq0$ with significance level $0.05$. If the test is not passed, the route is labeled as unstable. 

\begin{figure}[t]
\centering
{\includegraphics{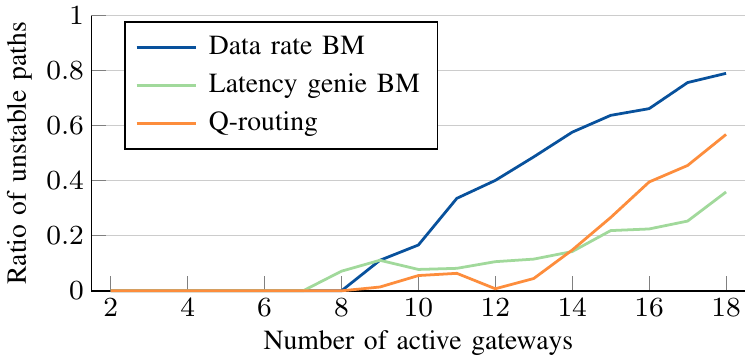}}
\caption{Ratio of unstable paths, for which $\text{H}_0\!: \beta_1 \leq 0$ is rejected for a significance level $0.05$, for the three schemes.}
\label{fig:congestion} \vspace{-0.2cm}
\end{figure}

Fig.~\ref{fig:congestion} shows the ratio of unstable paths: the source-destination pairs for which the null hypothesis $\text{H}_0\!: \beta_1\leq 0$ is rejected and, hence, it is concluded that the slope $\beta_1$ is positive. The ratio of unstable paths increases rapidly for the data rate \gls{bm} from $9$ active gateways. This is because, even though a path with high data rates is selected, the metric cannot adapt the path to the presence of congestion. In contrast, the ratio of unstable paths for the latency genie \gls{bm} increases at a slower pace than for the data rate \gls{bm} despite the initial increase occurs with $8$ active gateways. Finally, the initial increase in the ratio of unstable paths with Q-routing is slower than with both of the other metrics. It is only with more than $14$ gateways that Q-routing presents more unstable paths, but the ratio of unstable paths for all schemes is above $0.1$, which clearly makes the network unstable at this point. We conclude that the Q-routing metric is more robust to congestion than both the data rate and the latency genie \glspl{bm} in every case where the network is still operative. The fact that the ratio of unstable paths with the latency genie \gls{bm} increases with less active gateways than with the data rate \gls{bm} and with Q-routing illustrates that this is not an optimal routing approach even though it has instantaneous knowledge about the state of the queues. We have further looked at the individual paths with congestion in the cases with $8$ and $9$ active gateways and it turns out that all the unstable paths share similar characteristics: 1) the distance between the gateways is large, 2) go from south to north, and 3) most of them have the same destination. Because the latency genie \gls{bm} scheme selects the route for the packets from the source based on the instantaneous state of the queues, it cannot consider the state of the queues when the packets arrive to the corresponding links but only when these are about to leave the source. If the distance is long, the queue of a distant satellite might be empty at the time the packet is transmitted from the source, but increase significantly before the packet arrives. 
 This clearly shows that source routing is sub-optimal, even under the assumption of global and instantaneous knowledge. 
 In contrast, there are no unstable paths with Q-routing and $8$ active gateways. For the case of $9$ active gateways, the only unstable path with Q-routing is from Inuvik, Canada to Cordoba, Argentina. Hence, we conclude that Q-routing supports a higher traffic load as compared to the benchmarks. 

\begin{figure}[t]
    \centering
    \includegraphics{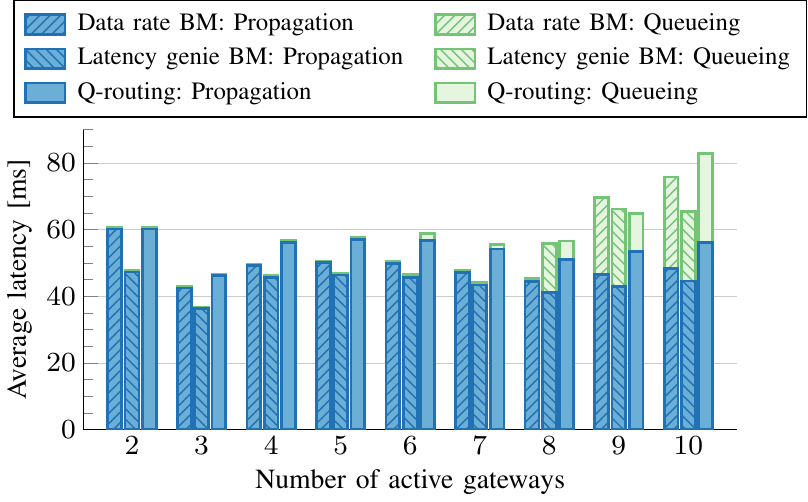}
    \caption{Average E2E propagation and queueing latency for the three schemes as a function of the number of active gateways.}
    \label{fig:avg_latency} \vspace{-0.5cm}
\end{figure}

\begin{figure}[t]
\centering
{\includegraphics[width=3.5in]{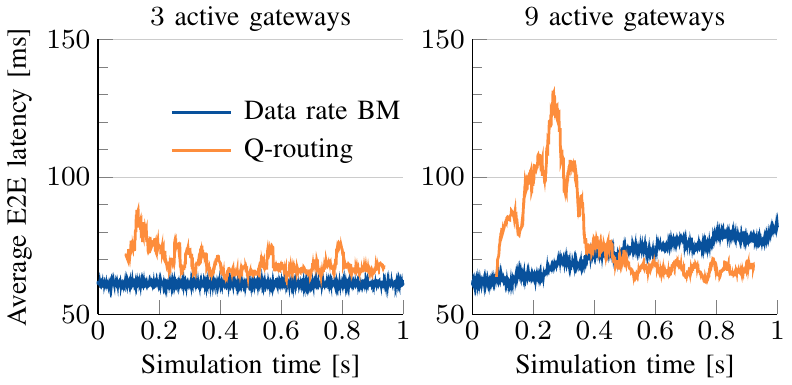}}
\caption{E2E latency versus simulation time. Comparison of Q-routing with the data rate \gls{bm}.}
\label{fig:timeevolution}
\vspace{-0.5cm}
\end{figure}

Once the system congestion is characterized, we analyze the different contributors to the E2E latency in Fig.~\ref{fig:avg_latency}. The transmission latency in all cases is below $0.72$~ms. Therefore, it is negligible compared to the propagation and queueing latency and not depicted. We plot the cases between $2$ and $10$ active gateways, for which the ratio of unstable paths is below $0.2$. As it can be seen, the latency genie \gls{bm} leads to the minimum propagation latency in all cases. Nevertheless, its queueing latency increases significantly with $|\mathbb{G}| \geq 8$. The propagation latency with Q-routing is a bit larger than the other two in all cases. There are two reasons for this: (1) we include the exploration stage at the beginning of the simulation when the satellites are mostly trying random paths (see Fig.~\ref{fig:timeevolution}); (2) we use a simple encoding of the status of the link to limit the size of the state space. In the future, we will explore the use of \gls{drl} which allows enlarging the state space, although at the expenses of a higher computation complexity. 

Finally, Fig.~\ref{fig:timeevolution} shows the time-evolution of the E2E latency of Q-routing and compares it to the data rate \gls{bm} with 3 and 9 gateways. It is observed that the Q-routing latency is larger in the beginning, when the system explores new paths, and rapidly decreases to a stable value and exploitation-intensive operation. The \gls{bm} keeps a low value with $|\mathbb{G}|=3$, but experiences a linear increase with $|\mathbb{G}|=9$ which reflects the system congestion.

\section{Conclusions} \label{sec:conclusions}
We propose a distributed Q-learning solution for E2E routing in \gls{lsatc} that leverages on the knowledge of the neighbours and the correlation of the routing decisions of each node. The results are compared to centralized benchmark solutions based on shortest path and full knowledge of the network. The Q-routing algorithm is demonstrated to be comparable in terms of E2E delay, while it supports a higher traffic load and it is simple enough to be implemented in practice, which is a key advantage in satellite technology. Future work will look at the extension of the state space to \gls{drl} and the evaluation in scenarios with heterogeneous QoS requirements and policies.

\bibliographystyle{IEEEtran}
\bibliography{ref}
\end{document}